\documentclass[conference]{IEEEtran}
\IEEEoverridecommandlockouts
\usepackage{cite}
\usepackage{amsmath,amssymb,amsfonts}
\usepackage{algorithmic}
\usepackage{graphicx}
\usepackage{textcomp}
\usepackage{xcolor}
\usepackage{balance}
\usepackage{hyperref}
\def\BibTeX{{\rm B\kern-.05em{\sc i\kern-.025em b}\kern-.08em
    T\kern-.1667em\lower.7ex\hbox{E}\kern-.125emX}}

\begin{document}
\title{The BET project: Behavior-enabled IoT}

%\thanks{This work is supported by the BeT project funded by MUR}

\author{\IEEEauthorblockN{ Henry Muccini}
\IEEEauthorblockA{\textit{DISIM Department} \\
\textit{University of L'Aquila and CINI} \\
L'Aquila, Italy \\
henry.muccini@univaq.it}
\and
\IEEEauthorblockN{ Barbara Russo}
\IEEEauthorblockA{\textit{Faculty of Engineering} \\
\textit{Free University of Bozen-Bolzano and CINI}\\
Bolzano, Italy \\
brusso@unibz.it}
\and
\IEEEauthorblockN{Eugenio Zimeo}
\IEEEauthorblockA{\textit{Dept. of Engineering } \\
\textit{University of Sannio and CINI}\\
Benevento, Italy \\
zimeo@unisannio.it}}
%\date{June 2023}

\maketitle

\begin{abstract}
IoT is changing the way Internet is used due to the availability of a large amount of data timely collected from every-day life objects. Designing applications in this new scenario poses new challenges.
This extended abstract discusses them and presents the objective of the BeT project whose main aim is to introduce a reference architecture, a conceptual framework, and related techniques to design behavior-enabled IoT systems and applications.
\end{abstract}


\section{Introduction}
\label{introduction}
The increasing adoption of the Internet of Things (IoT) combined with the rapid expansion of the edge-to-cloud computing
continuum, the widespread adoption of AI technologies, as well as the tighter participation of human beings and their personal
devices, are pushing toward an expanded notion of the Internet where {\em humans become more central than ever}. 

%The pandemic circumstances are responsible for a huge acceleration of such a transition towards a digital society in which humans are more and more active agents rather than just service consumers. This is known as “people centricity”: humans affect the digital world and the digital world affects in turn humans, causing nontrivial cause-effect relations and emergent behaviors. 

%Because of this transition, businesses are currently rethinking how they communicate and interact with customers.
Gartner in its “Gartner Top Strategic Technology Trends for 2021” \cite{gartner} recognized the Internet of Behaviors (or simply, IoB) as a strategic technology trend needed by resilient businesses to overcome the current economic state of the world. 
IoB collects
human behaviors (through an underline IoT network or social networks), analyzes them from a behavioral psychology standpoint, and uses them to influence users’ behaviors. 
%Companies such as Amazon, Netflix, and Booking.com are making use of the IoB; the sales industry, insurance companies, and the healthcare domain are using the IoB in order to adjust recommendations according to human behavior.
%The potential of IoB is enormous since both IoT, AI, and human behaviors are taken into account when making decisions. 
 
So far, IoB
applications are realized by unsystematically building layers on top of traditional IoT reference architectures. 
This process is
error-prone and based on personal-, rather than, best- practices. 

The BeT (Behavior-enabled IoT) project aims at enhancing the way IoB systems are designed and implemented. It defines 
a reference architecture and a set of
well-defined and validated methods and patterns for systematically engineering the IoB. 
.

\section{BeT Model and challenges}
BeT foresees a network of humans and computational elements (e.g., IoT devices, networks, data, and services) that, through AI and machine learning algorithms, may give rise to a network of cooperating human-digital entities. 

This view goes beyond the typical scenario where humans are mobile agents immersed in a static IoT space \cite{9779704}. Both humans and cyber-physical entities can move in a shared environment, each one impacting the behavior of the other.
This implies a \emph{bi-causal} connection between humans and software systems: on the one hand, humans’ (and communities’) behaviors may be collected and analyzed in order to influence the behavior of other people with suggestions and decisions that may maximize their expectations and their Quality of Experience (QoE); on the other hand, software behavior itself may be adapted to follow human expectations and actions, so to guarantee Quality of Service (QoS). 

In the BeT vision, human (and communities) desires and recurring behaviors have to be inferred, modeled, and validated. 
Moreover, AI-enabled components support the inference of human and system behaviors, based on recurring behavioral patterns \cite{DBLP:conf/hicss/KjaergaardMAM23}.

The BeT proposal tries to address a set of challenges:

\emph{Human-System behavior Interaction and its Interpretation:} Human agents in BeT are a key part of the environment and/or the system itself. To provide strong guarantees for QoS and QoE in these systems, we need to address the limitations of current human modeling techniques and deal with the inherent variability and uncertainty in their behavior. Finding the right fidelity and keeping these models alive in production is fundamental to avoiding highly over-approximate knowledge that could lead in turn to bad QoE.

\emph{Human-System bi-causal Quality connection effects:} Dynamic changes in the system itself and the surroundings can be interleaved giving rise to nontrivial unforeseen cause-effect mutual relations between system and human behaviors. Understanding the mutual influence between QoS and QoE represents an emerging problem in IoB systems.

\emph{Cooperative adaptation between humans and AI-enabled IoT:} Humans and AI-enabled IoT (AIoT) systems, while interacting and collaborating, shall adapt to each other in order to maximize QoE and QoS.  While self-adaptation techniques are well studied in IoT applications, the joint human-AIoT adaptation is still to be studied.

\section{BeT Architecture and goals}
A reference architecture is a software architecture template defining the structures, elements and relations to realize concrete architectures in specific domains or families of software systems. 
While IoT reference architectures have been extensively discussed in the literature, and some recent work discusses preliminary AIoT reference architectures, BeT aims at defining its own reference architecture for IoB applications. 

\begin{figure*}[h]
    \centering
    \includegraphics[scale=0.6]{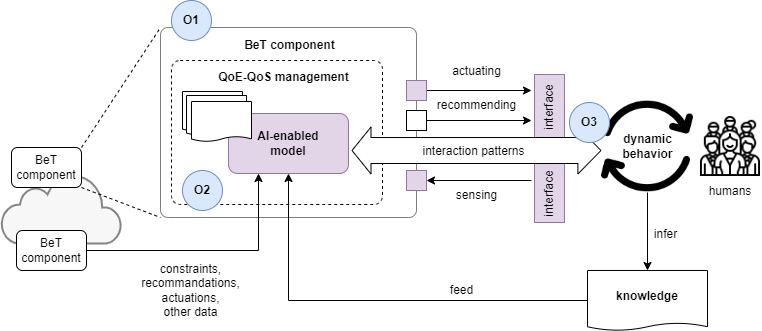}
    \caption{BeT high-level reference architecture }
    \label{fig:architecture}
\end{figure*}

The BeT reference architecture includes AI
components to analyze and predict human behaviors, as well as Quality of Service (QoS) and Quality of Experience (QoE) metrics to
maximize the system-specific and human-perceived qualities. 
BeT provides methods to model human behavior patterns, as well as means to detect QoE-QoS antipatterns. 
A bi-causal connection between humans and systems is defined, in order to maximize QoE and QoS. Self-adaptiveness embedded in BeT supports the human and system behavior run-time autonomous adaptation to users’ and system’s needs. %The BeT will be validated on a set of real use cases owned by the partners

BeT foresees a network of humans and computational elements (e.g., IoT devices, networks, data, and services) that, through AI and machine learning algorithms, may give rise to a network of cooperating human-digital entities. 

This view goes beyond the typical scenario where humans are mobile agents immersed in a static IoT space.
Both humans and cyber-physical entities can move in a shared environment, each one impacting the behavior of the other.
This implies a  \emph{bi-causal connection} between humans and software systems: on the one hand, humans’ (and communities’) behaviors may be collected and analyzed in order to influence the behavior of other people with suggestions and decisions that may maximize their expectations and their Quality of Experience (QoE); on the other hand, software behavior itself may be adapted to follow human expectations and actions, so to guarantee Quality of Service (QoS). 
In the BeT vision, human (and communities) desires and recurring behaviors have to be inferred, modeled, and validated. 
Moreover, AI-enabled components support the inference of human and system behaviors, based on recurring behavioral patterns.

BeT aims at supporting the design of behavior enabled IoT systems that satisfy quality of service and human (as an individual or group of people) expectations at the same time. For example, in the evacuation context, people could rapidly reach the exits (maximizing the QoE) but this can create data congestion in the IoT (not optimized QoS); a car trip planner could exploit AI-enabled models to suggest uncongested, shortest paths but this could lead to an unsatisfactory response time due to a more complex computation; an infrastructure autoscaler continuously follows the traffic of the incoming requests to allocate resources but this could lead to bad QoE due to continuous reallocation that could be avoided with a better knowledge of users’ behavior. Therefore, QoE and QoS and their possible interrelations assume an important role. 

BeT aims at expressing behaviors originated by both human and system changes and self-adapt to such bi-causal behavior stimuli while maintaining a trade-off between QoE and QoS. To this aim, BeT will detect and encode (anti)patterns as solutions to recurring human-system interactions that have (negative)positive effects on the design and the trade-off between QoE and QoS in an AIoT system. Real use cases provided by the research units will show the (anti)patterns “in action”. These (anti)patterns and the related use cases can be used by architects or behavior scientists to compare different BeT instantiations as well as to implement new AIoT systems that aim at incorporating human-system interactions and their effects on QoE and QoS by design.

\section{Conclusions}
BeT, a PRIN 2022 accepted project, will investigate methods, patterns, and a reference architecture for systematically engineering the IoB. The project is expected to start in September 2023. Anybody interested to collaborate is welcome to contact the authors.

\balance

\bibliographystyle{IEEEtran}
\bibliography{references}
\end{document}